\documentclass[conference]{IEEEtran}
\usepackage{amsfonts}
\usepackage{amsmath}
\usepackage{amsthm}
\usepackage{verbatim}
\usepackage{setspace}
\usepackage[caption=false,font=footnotesize]{subfig}

\newtheorem{theorem}{Theorem}
\newtheorem{lemma}{Lemma}

\newtheorem{corollary}{Corollary}

\newtheorem{remark}{Remark}

\usepackage{cite}

\ifCLASSINFOpdf
   \usepackage[pdftex]{graphicx}
\else
\fi

\def\bb0{{\mathbb{0}}}

\newcommand{\mytilde}{\raise.17ex\hbox{$\scriptstyle\mathtt{\sim}$}}

\def\bb{{\mathbf{b}}}

\def\b0{{\mathbf{0}}}

\def\sf0{{\mathsf{0}}}

\usepackage{algorithm}
\usepackage{algpseudocode}
\usepackage{epstopdf}
\usepackage{balance}
\usepackage{bbm}
\usepackage{todo}
\IEEEoverridecommandlockouts

\newcounter{storeeqcounter}
\newcounter{tempeqcounter}

\begin{document}
%
\title{A Stochastic Geometry Approach to Doppler Characterization in a LEO Satellite Network}


\author{Talha Ahmed Khan${}^{*}$ and Mehrnaz Afshang${}^{*}$
\thanks{The authors are with Ericsson Research, Santa Clara, USA (Email: \{talha.khan, mehrnaz.afshang\}@ericsson.com).}
}
\maketitle
\begin{abstract}
A Non-terrestrial Network (NTN) comprising Low Earth Orbit (LEO) satellites can enable connectivity to under-served areas, thus complementing existing telecom networks. The high-speed satellite motion poses several challenges at the physical layer such as large Doppler frequency shifts. 
In this paper, an analytical framework is developed for statistical characterization of  Doppler shift in an NTN where LEO satellites provide communication services to terrestrial users.
Using tools from stochastic geometry, the users within a cell are grouped into disjoint clusters 
to limit the differential Doppler across users. 
Under some simplifying assumptions, the cumulative distribution function (CDF) and the probability density function  are derived for the Doppler shift magnitude at a random user within a cluster. The CDFs are also provided for the minimum and the maximum Doppler shift magnitude within a cluster.   
Leveraging the analytical results, the interplay between key system parameters such as the cluster size and satellite altitude is examined. Numerical results validate the insights obtained from the analysis.
\end{abstract}

%
\IEEEpeerreviewmaketitle

\section{Introduction}
Satellite communications can provide connectivity to under-served regions, thus augmenting existing telecommunications ecosystem\cite{lin20195g,guidotti2019architectures}. 
A constellation of Low-Earth Orbit (LEO) satellites is an essential ingredient of a satellite-enabled telecommunications network. 
Despite its economic advantages, satellite communications in LEO orbits poses several 
technical challenges owing to high-speed satellite motion.
The chief concern at the physical layer is to cope with substantially high Doppler frequency shifts. One solution is to leverage frequency compensation techniques to at least partially eliminate the Doppler shift. Unfortunately, the residual Doppler frequency shift could still be large enough to compromise the subcarrier orthogonality in a multicarrier system. For system design, it will be useful to have a statistical characterization of the Doppler shifts in the service area of a LEO satellite. 

\subsection{Related work} 
Mobile satellite communications was an active research area in the 90's thanks to commercial ventures such as Iridium and Globalstar \cite{lin20195g,guidotti2019architectures}. There has been extensive work on the performance characterization of satellite communication systems \cite{ganz1994performance,ali1998doppler,ali1997doppler,li2002analytical}.
In \cite{ganz1994performance}, a system-level capacity analysis of a LEO satellite network was presented. The average number of beam-to-beam and inter-satellite handovers, the channel occupancy distribution and the average call drop probability were examined as a function of the network constellation, the satellite speed, the cell size, and the average transaction duration. 
In \cite{ali1998doppler}, an analytical treatment for the Doppler shift in a LEO satellite link was provided. Using spherical geometry, an accurate expression for the Doppler shift was derived for a point on the earth's surface. The satellite visibility duration was shown to be a function of the maximum elevation angle. Numerical results validated the analytical expressions for circular-orbit satellites with altitude up to 10,000 km. In \cite{ali1997doppler}, a Doppler-based multiple access protocol for a bent-pipe LEO satellite system was proposed. In the satellite visibility footprint, regions of eligibility were configured based on the parameters of the Doppler-time curves at the users. By allowing access only to users located in a region of eligibility, the uplink flow control was shown to improve. Leveraging the work in \cite{ali1998doppler}, an accurate analytical model to predict the elevation angle distribution for a LEO constellation was proposed in \cite{li2002analytical}. 

The recent momentum generated by the confluence of the satellite and telecom industries has rekindled the interest in mobile satellite systems\cite{lin20195g,guidotti2019architectures}. 
In 3GPP Release 16, there is an ongoing study item \cite{3GPPTR} on Non-terrestrial Networks (NTN) for New Radio (NR) with a long-term goal to integrate satellite communications into 5G systems. With this motivation, satellite communications has been revisited in the context of 5G\cite{guidotti2017satellite,maxdoppler2018NBIoT,kodheli2019journal,guidotti2019architectures}. 
In \cite{guidotti2017satellite}, the impacts of delay and Doppler shift on a satellite-enabled Long-term Evolution (LTE) system were discussed. A network architecture was considered where a LEO constellation forms the backhaul infrastructure for terrestrial relay nodes serving LTE users. Hybrid Automatic Repeat reQuest (HARQ) procedure and handover were identified as the key challenges in deploying the proposed architecture.
In \cite{maxdoppler2018NBIoT,kodheli2019journal}, leveraging the work in \cite{ali1998doppler}, the maximum differential Doppler in a LEO-based Narrowband-Internet of Things (NB-IoT) cell was derived. An uplink resource allocation scheme was proposed to reduce the maximum differential Doppler within a cell. Simulation results validated that the proposed technique reduces the differential Doppler to levels tolerated by the existing NB-IoT protocol.
In \cite{guidotti2019architectures}, some of the potential challenges facing 5G satellite systems were outlined. 
With NR standard as the baseline, several solutions were discussed for adapting physical and control layer procedures such as random access, timing advance, and HARQ for various NTN operating scenarios.

\subsection{Contributions}
In this paper, we leverage stochastic geometry to obtain a statistical characterization of the Doppler shifts in an NTN where LEO satellites provide communication services to stationary users.
To limit the differential Doppler shift, we group the terrestrial users using a cluster point process. We derive the cumulative distribution function (CDF) and the probability density function (PDF) for the Doppler shift magnitude at a random user within a cluster. We also provide analytic expressions for the distribution of the minimum and the maximum Doppler shift magnitude in a cluster. The analytical framework exposes the interplay between key system parameters such as cluster size, satellite altitude, number of users, carrier frequency, and satellite speed. Numerical simulations validate the accuracy of the technical results. 
We note that prior work \cite{ali1998doppler,guidotti2017satellite,maxdoppler2018NBIoT,kodheli2019journal} on Doppler shift in satellite scenarios deals with a deterministic set of user locations. Our key contribution is to statistically characterize the  Doppler shifts with clustering while accounting for the topological randomness in network geometry. The presented analytical results reveal several design insights for a LEO satellite communications system.

The rest of this paper is organized as follows. The system model is described in Section \ref{secSys}. The Doppler magnitude distribution is presented in Section \ref{secAnl}. Simulation results are provided in Section \ref{secSim}. The paper is concluded in Section \ref{secConc}.

\section{System Model}\label{secSys}

In this paper, we consider the scenario where a constellation of circular-orbit LEO satellites serve stationary user equipments (UEs) located 
on the earth's surface. Each satellite can form multiple spotbeams within its coverage area. We call the footprint of a satellite spotbeam a cell. This paper provides Doppler characterization within one such cell. 
The satellite can be transparent or regenerative with either earth-fixed or moving-earth beams. All signals transmitted or received by a satellite on service links usually go through the same feeder link. Therefore, we focus on the service link without loss of generality. The signal transmitted or received by the satellite undergoes a Doppler frequency shift due to the relative motion between the satellite and UEs.
The relative velocity between the satellite and the UE is a function of the satellite elevation angle and satellite altitude. The Doppler shift can vary considerably among the UEs within a cell due to differences in the UE elevation angles to the satellite. Using frequency compensation techniques relative to a reference point within the cell, it is possible to partially eliminate the Doppler shift common to all the UEs. While the reference point sees no Doppler shift, this leaves other locations with a residual Doppler shift whose magnitude depends on the proximity of the UE location to the reference point. The difference in the residual Doppler shifts between any two UEs causes uplink interference due to loss of subcarrier orthogonality. The differential Doppler shift is typically higher for UEs located further away from each other. This motivates scheduling UEs such that those located sufficiently apart are not served concurrently to avoid large differential Doppler shifts. This can help curtail the maximum differential Doppler shift in a cell.
\begin{figure}
	\includegraphics[width=\columnwidth]{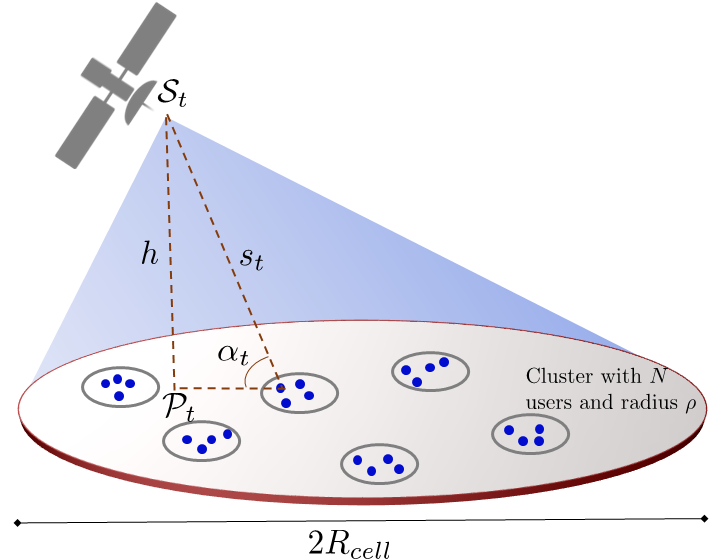}
	\caption{System model.}
	\label{fig: sys model}
\end{figure}
We consider the scenario where the UEs within each cell are grouped into disjoint clusters based on their locations. To reduce the impact of differential Doppler, the clusters within a given cell are served using time division multiple access (TDMA)
on orthogonal time resources. The UEs within each cluster can be scheduled concurrently.  
To simplify the analysis, we approximate the cell formed by a spotbeam on the earth's surface 
 by a circular region of radius $R_{\text{cell}}$ on a plane $\mathbb{P}$.  
We consider a 3D coordinate system where the origin $\mathcal{O}=\{0,0,0\}$ lies on the plane $\mathbb{P}$ containing the cell and the UEs. We  denote the satellite location at time $t$ as $\mathcal{S}_t=\{s_x(t),s_y(t),s_z(t)\}$ relative to the origin. We further define $\mathcal{P}_t=\{s_x(t),s_y(t),0\}$ as the projection of $\mathcal{S}_t$ on $\mathbb{P}$. 
We use $\mathcal{C}_t$ for the cluster served by the satellite at time $t$. The network geometry for the considered scenario is illustrated in Fig. \ref{fig: sys model}, where $s_{t}$ is the slant range and $\alpha_{t}$ is the elevation angle for a particular UE in $\mathcal{C}_t$ and $h$ is the altitude of the satellite.

We assume the UEs are spatially distributed on $\mathbb{P}$ according to a cluster process where the parent points model the locations of virtual cluster heads while the daughter points model the locations of the UEs within a cluster. Formally, the virtual cluster heads are drawn independently from a homogeneous point process $\Phi_{c}$ with density $\lambda_c$. We assume that there are $N$ daughter points
 per cluster such that the overall UE density is $N\lambda_c$. 
The suitability of a Matern cluster process (MCP) for modeling many practical user distributions is well-established\cite{3GPPHetNetSahaAfsShi2018}.
 We model the daughter point process using a variant of MCP where the number of points per cluster is fixed.  
\begin{align}
f_{\text{MCP}}(x)=\begin{cases}
\frac{1}{\pi\rho^2}, \qquad\|x\|\leq \rho \\
0, \quad\qquad \text{else}.\\
\end{cases}
\end{align}
As $\rho$ ($0 <\rho\leq R_{\text{cell}}$) describes the cluster size, $\rho=R_{\text{cell}}$ means that the cluster is as large as the cell itself. We use it as a proxy for the case where the UEs within the cell are not clustered and can be scheduled concurrently.
We note that it is possible that a cluster head within a certain cell area has one or more UEs physically located outside it. 
We assume that all UEs associated with a cluster are served by the same cell that the cluster head belongs to. 

In the ensuing analysis, we characterize the Doppler shift distribution within a cluster at a given point in time. 
We use subscript M for MCP 
in the notation for the analytical expressions. We define $\mathcal{B}(u,v)$ as a circle of radius $v$ centered at a point $u$ on $\mathbb{P}$.

\section{Analytical Results}\label{secAnl}
In this section, we derive the distributions of the magnitudes of Doppler shift, the minimum Doppler shift and the maximum Doppler shift within a cluster. A key assumption is approximating UE locations to be on a plane instead of a spherical surface. 
This allows leveraging the existing results for distance distribution of a random point inside a circle, and also simplifies the calculation of UE location relative to the sub-satellite point. 
The satellite may or may not be located over the cluster. To simplify analysis, we consider two cases depending on whether the sub-satellite point $\mathcal{P}$ lies within or outside the circle $\mathcal{B}(\mathcal{C},\rho)$ centered at the cluster center $\mathcal{C}$. 
\subsection{Doppler shift formula}
We use the following analytical approximation for the Doppler shift at a stationary UE located on the earth surface. We define $\omega_F$ as the satellite angular velocity (radians/sec) in ECF (Earth centered fixed) frame. For LEO satellites, as shown in \cite{ali1998doppler}, it can be tightly approximated  as $\omega_F\approx\omega_s+\omega_E\cos\left[\theta_i\right]$ where $\omega_s$ and $\omega_E$ respectively denote the angular velocities of the satellite and the earth in the ECI (Earth centered inertial) frame, and $\theta_i$ is the orbital inclination (radians) of the satellite relative to the equatorial plane.

\begin{lemma}\normalfont\label{lem:doppler formula}
		Let $\alpha_t\in\left[0,\frac{\pi}{2}\right]$ be the elevation angle from a stationary UE on the earth's surface to a circular orbit LEO satellite at time $t$. The UE observes the maximum elevation angle $\alpha_{\max}$ at time $t=t_{\alpha_{\max}}$. The Doppler shift $\chi_t$ due to satellite motion can be analytically expressed as
	\begin{align}\label{eq: doppler}
	\chi_t= -\frac{f_c~r_E~\dot{\gamma_t}~\cos[\alpha_t]}{c},
	\end{align}
	where $f_c$ is the carrier frequency, $r_E$ is the earth radius, $c$ is the speed of light,  
	\begin{align}\label{eq: derivative gamma_t}
	\dot{\gamma_t } =\frac{\omega_F~ \Theta\left[\alpha_{\max}\right]\sin\left[(t-t_{\alpha_{\max}})\omega_F\right]}{\sqrt{1-\Theta^2\left[\alpha_{\max}\right]\cos^2\left[(t-t_{\alpha_{\max}})\omega_F\right]}}
	\end{align}
	is the rate of change of central angle between the sub-satellite point and the UE location, $r_o = r_E+h$ is the orbital radius, and
	\begin{align}\label{eq: function theta max}
	\Theta\left[\alpha_{\max}\right]=\cos\left[\cos^{-1}\left[\frac{r_E}{r_o}\cos\left[\alpha_{\max}\right]\right] - \alpha_{\max}\right].
	\end{align}
	\begin{proof} This lemma is obtained by algebraic manipulations of the expressions derived in \cite{ali1998doppler}. The key underlying assumptions in \cite{ali1998doppler} were to approximate the satellite trajectory relative to the earth by a great circle arc and the satellite angular velocity in the ECF frame by a  constant. In \cite[Eq. (5)]{ali1998doppler}, the instantaneous Doppler shift is given by
		\begin{align}\label{eq: doppler formula}
		\chi_t&=-\frac{f_c}{c}\frac{r_E~ r_o~\omega_F\sin\left[(t-t_{\alpha_{\max}})\omega_F\right]\Theta\left[\alpha_{\max}\right]}{s_t},
		\end{align}
	where $\Theta\left[\alpha_{\max}\right]$ is defined in (\ref{eq: function theta max}) and follows from \cite[Eq. (4)]{ali1998doppler}. Invoking \cite[Eq. (1) and (4)]{ali1998doppler}, the slant range $s_t$ is expressed in terms of the maximum elevation angle $\alpha_{\max}$ as   
		\begin{align} \label{eq: slant range}
		s_t=\sqrt{{r_E}^2+r_o^2-2r_or_E\cos\left[(t-t_{\alpha_{\max}})\omega_F\right]\Theta\left[\alpha_{\max}\right]}.
		\end{align}
By differentiating \cite[Eq. (2)]{ali1998doppler}, we obtain the relation 
\begin{align}\label{eq: differentiate 2}
\sin\left[\gamma_t\right]\dot{\gamma_t}=\omega_F\Theta\left[\alpha_{\max}\right]\sin\left[ (t-t_{\alpha_{\max}})\omega_F\right],
\end{align}
where $\gamma_t$ is the central angle between the subsatellite point and the UE location at time $t$, and its time derivative $\dot{\gamma_t}$ given in (\ref{eq: derivative gamma_t}) also follows from \cite[Eq. (2)]{ali1998doppler}.
From \cite[Eq. 3-4]{pritchard1993satellite}, the following relation is obtained 
		\begin{align}\label{eq: ele angle}
		\cos\left[\alpha_t\right]=\frac{r_{o}\sin\left[\gamma_t\right]}{s_t}.
		\end{align}
Finally, (\ref{eq: doppler}) follows by substituting (\ref{eq: function theta max}), (\ref{eq: differentiate 2}) and (\ref{eq: ele angle}) in (\ref{eq: doppler formula}).
\end{proof}
\end{lemma}

\begin{corollary}\normalfont\label{cor: simple doppler}
For the special case where the UE lies on the ground track such that $\alpha_{\max}=\frac{\pi}{2}$ and $\dot{\gamma_t}=\omega_F$, the Doppler shift formula in (\ref{eq: doppler}) simplifies to
\begin{align}\label{eq: doppler formula special}
\chi_t&=-\frac{f_c~r_E~ \omega_F~ \cos\left[\alpha_t\right]}{c}.
\end{align} 
\end{corollary}

The following result establishes that the expression in (\ref{eq: doppler}) is in fact a bound on the Doppler shift experienced by UEs lying outside the orbital plane. 

\begin{lemma}\normalfont\label{lem:doppler formula bound}
With other parameters fixed, the Doppler shift
at a UE with maximum elevation angle $0<\alpha_{\max}<\frac{\pi}{2}$ given in (\ref{eq: doppler}) is bounded by the expression in (\ref{eq: doppler formula special}).  
\begin{proof}
We now prove that the Doppler shift magnitude $|\chi_t|$ obtained from (\ref{eq: doppler formula special}) is an upper bound on that given in (\ref{eq: doppler}) for $\alpha_{\max}<\frac{\pi}{2}$. The proof is obtained by the following observations.
First, we note that for a fixed $r_o$, $\Theta\left[\alpha_{\max}\right]$ in (\ref{eq: function theta max}) is an increasing function of $\alpha_{\max}$. For $\alpha_{\max}\in[0,\frac{\pi}{2}]$, the function range is $\Theta[\alpha_{\max}]\in[\frac{r_E}{r_o},1]$ where the ratio $\frac{r_E}{r_o}=\frac{r_E}{r_E+h}<1$. Therefore, $\Theta\left[\alpha_{\max}\right]<1$ for $\alpha_{\max}\neq\frac{\pi}{2}$.
Second, we further note that $|\dot{\gamma_t}|<\omega_F$ for $\Theta\left[\alpha_{\max}\right]<1$. This follows by applying the following inequality on (\ref{eq: derivative gamma_t}): $\frac{|\sin\left[x\right]|}{\sqrt{1-y^2\cos^2\left[x\right]}} < 1$ for $0<y<1$. These observations complete the proof. 
	\end{proof}
\end{lemma}

\begin{remark}\normalfont
	Let us quantify the tightness of the bound proposed in Lemma \ref{lem:doppler formula bound}. We note that the bound is exact for $\Theta[\alpha_{\max}]=1$. Let $\epsilon\in(0,1]$ be the normalized error magnitude between the exact Doppler in (\ref{eq: doppler}) and the bound in (\ref{eq: doppler formula special}), i.e., $\epsilon \chi_t$ is the absolute gap when $\chi_t$ is as given in (\ref{eq: doppler formula special}). Then, for $\Theta[\alpha_{\max}]<1$, the bound is $\epsilon$-accurate at the time instants 
	\begin{align}\label{eq: error}
	t_\epsilon=t_{\alpha_{\max}}+\frac{1}{\omega_F}\cos^{-1}\left[\pm\sqrt{\frac{1-\frac{(1-\epsilon)^2}{\Theta^2[\alpha_{\max}]}}{1-(1-\epsilon)^2}}\right].
	\end{align}
In other words, the bound is at least $\epsilon$-accurate for the time range defined by (\ref{eq: error}). Therefore, given the satellite and cluster parameters, we can obtain the time range (or equivalently the set of satellite locations) for which the proposed analysis yields sufficiently accurate CDF. %
\end{remark}

\begin{remark}\normalfont \label{rem: approximation assumption} We note that different UEs within a cluster may have different maximum elevation angles. To simplify the analysis, we leverage Lemma \ref{lem:doppler formula bound} and use the expression in (\ref{eq: doppler formula special}) to characterize the Doppler shift at the UEs. This eliminates the dependency of the Doppler shift on the maximum elevation angle. We acknowledge our pessimistic approach since the bound over-estimates the Doppler shift magnitude for UEs outside the orbital plane. We noted in our experiments that this simplifying assumption still yields a reasonably accurate estimate of Doppler shift magnitude in the context of this paper.
\end{remark}
For a given UE location, the satellite elevation angle changes with time due to satellite motion. 
Since the satellite trajectory is predictable, the resulting Doppler shift at that location can be described analytically by a Doppler time curve. 
At a given time, however, the UEs within a cluster will typically experience different Doppler shifts.
Due to the underlying randomness in UE locations, the resulting elevation angles $\{\alpha_{t,u_i}\}_{i=1}^{N}$ and the Doppler shifts $\{\chi_{t,u_i}\}_{i=1}^{N}$ for the $N$ in-cluster UEs at time $t$ will also be random. We now characterize the statistical distribution of the random variable $|\chi_{t}|$, i.e., the Doppler shift magnitude at a random UE inside a given cluster.

\subsection{Doppler characterization for MCP}\label{secMCP}
In this subsection, we provide an analytical treatment of the Doppler shift for an MCP. The following distance distribution lays the groundwork for the ensuing analysis.

\begin{lemma}\normalfont\label{lem:cdf thomas distance}
	The CDF of the distance between a random point and a fixed point where the former is located uniformly in a circle of radius $R$, while the latter is 
	located arbitrarily on the same plane as the circle is 
given by (\ref{eq: distance cdf}) on the top of the next page,
\setcounter{storeeqcounter}%
{\value{equation}}%
\begin{figure*}[!t]
	\normalsize
	\setcounter{tempeqcounter}{\value{equation}} 
	\begin{align}\label{eq: distance cdf}		\setcounter{equation}{\value{storeeqcounter}} 
	F_{\text{M}}(r,R|\hat{R}) = \begin{cases}
	\frac{r^2}{R^2}, {0 \leq}  r\leq R-\hat{R}\\
	\frac{r^2}{\pi R^2} \left(\theta^*-\frac{1}{2}\sin(2\theta^*)\right) +\frac{1}{\pi} \left(\phi^*-\frac{1}{2}\sin(2\phi^*)\right),   |R-\hat{R}| <  r \leq R+\hat{R},\\
	\end{cases}
	\end{align} 		
	\setcounter{equation}{\value{tempeqcounter}} 
	\hrulefill
	\vspace*{4pt}
\end{figure*}
\addtocounter{equation}{1}%
where $\hat{R}$ is the distance of the fixed point from the center of the circle, 
 \begin{align}
 \theta^*={\cos}^{-1}\left[\frac{r^2+\hat{R}^2-R^2}{2\hat{R}r}\right]
 \end{align}
  and
  \begin{align} \phi^*={\cos}^{-1}\left[\frac{R^2+\hat{R}^2-r^2}{2\hat{R}R}\right].
  \end{align}
	\begin{proof}
	See \cite[Theorem 2.3.6]{mathai1999introduction}.
	\end{proof}
\end{lemma}

We note that the fixed point can be located inside or outside the circle.

\begin{lemma}\normalfont\label{lem: matern cdf}
	The PDF corresponding to the distance distribution in Lemma \ref{lem:cdf thomas distance} is given by
	\begin{align}\label{eq: dist pdf}
	f_{\text{M}}(r,R|\hat{R}) = \begin{cases}
	\frac{2r}{R^2},& {0 \leq}  r\leq R-\hat{R}\\
	\frac{2r \cos^{-1}\left[\frac{r^2+\hat{R}^2-R^2}{2\hat{R}r}\right]}{\pi R^2},  & |R-\hat{R}| <  r \leq R+\hat{R}.\\
	\end{cases}
	\end{align} 
	
	\begin{proof}
		See \cite[Theorem 2.3.6]{mathai1999introduction}.
	\end{proof}
\end{lemma}

The  following result provides the Doppler magnitude distribution within a cluster in an MCP given the satellite location at a certain time.

\begin{theorem}\normalfont\label{th: matern}
	For a satellite at altitude $h$ and slant range $s_{t}$ relative to {the cluster center}, the CDF $F_{\chi,\text{M}}^{t}(x)=\Pr\left[|\chi_{t}|\leq x \right]$ of the Doppler shift magnitude within a cluster of radius $\rho$ with a uniform UE distribution is given by
	\begin{align}\label{eq: thm doppler}
	F^{t}_{\chi,\text{M}}(x) = F_{{\text{M}}}\left(\frac{hx}{\sqrt{A^2-x^2}},\rho~\bigg| \hat{R}_{t}\right),
	\end{align}
	where $F_{\text{M}}(\cdot,\cdot|\cdot)$ follows from (\ref{eq: distance cdf}),  $A=\frac{f_c r_o \left(\omega_s+\omega_E\cos[\theta_i]\right) }{c}$ and $\hat{R}_{t}=\sqrt{s^2_{t}-h^2}$. 
	\begin{proof}
 Let us consider a random UE within a cluster of radius $\rho$ served by the satellite at time $t$. Then, the distance $z_t$ between the UE and the sub-satellite point at time $t$ is also a random variable. We invoke Lemma \ref{lem:cdf thomas distance}, set $R=\rho$ and $\hat{R}=\sqrt{s^2_{t}-h^2}$ to obtain the CDF for $z_t$. As evident from the geometry in Fig. \ref{fig: sys model}, 
the elevation angle for the UE can be expressed as $\cos[\alpha_t]=\frac{r_o \sin[z_t/r_E]}{s_t}\approx\frac{r_o z_t}{r_E s_t}=\frac{r_o z_t}{r_E\sqrt{h^2+z_t^2}}$. The next step is to substitute this expression in the Doppler formula in (\ref{eq: doppler formula special}), which we use for reasons mentioned in Remark \ref{rem: approximation assumption}. As a result, the Doppler shift $\chi_t$ can be expressed as a function of a random variable $z_t$ with a known distribution. The result in (\ref{eq: thm doppler}) follows from further algebraic manipulations.
	\end{proof} 
\end{theorem}
The time dependency in the conditional distribution in (\ref{eq: thm doppler}) follows from the fact that the satellite's location is a function of time.
We note that $A$ captures the effects of carrier frequency, satellite speed and inclination angle. It decreases with an increase in satellite altitude due to an underlying decrease in the satellite angular velocity.      
We further note that the CDF is independent of the number of UEs within the cluster. 
\begin{corollary}\normalfont\label{cor:pdf}
	For the CDF provided in Theorem \ref{th: matern}, the PDF is given by
\begin{align}\label{eq: thm doppler pdf}
f^{t}_{\chi,\text{M}}(x) = \frac{h
	A^2}{(A^2-x^2)^{3/2}}f_{{\text{M}}}\left(\frac{hx}{\sqrt{A^2-x^2}},\rho~\bigg| \hat{R}_{t}\right),
\end{align}
where $f_{\text{M}}(\cdot,\cdot|\cdot)$ follows from (\ref{eq: dist pdf}). 
\begin{proof}
	The proof is similar to that of Theorem 1.	
\end{proof} 
\end{corollary}
\begin{remark}\normalfont
	For a given satellite location, the Doppler shift magnitude at a randomly selected UE among the UEs in the cluster being served is 
	distributed according to (\ref{eq: thm doppler}). This follows by noting that the distance distribution in Lemma \ref{lem:cdf thomas distance} is i.i.d. across UEs within the cluster conditioned on satellite and cluster center locations, and the Doppler magnitude distribution in (\ref{eq: thm doppler}) is only a function of the distance distribution in Lemma \ref{lem:cdf thomas distance}. Therefore, the Doppler shift magnitude is i.i.d. among the in-cluster UEs.
\end{remark}
We now present some order statistics for the Doppler magnitude within a cluster.
\begin{corollary}\normalfont
	The CDF of the minimum Doppler magnitude $F_{\chi_{\min},\text{M}}^{t}(x)$ within a cluster with $N$ UEs is given by	$F_{\chi_{\min},\text{M}}^{t}(x)=1-\left(1-F^{t}_{\chi,\text{M}}(x)\right)^N$, where $F^{t}_{\chi,\text{M}}(x)$ follows from (\ref{eq: thm doppler}). The corresponding PDF is given by
	$f_{\chi_{\min},\text{M}}^{t}(x)=N \left(1-F^{t}_{\chi,\text{M}}(x)\right)^{N-1}f^{t}_{\chi,\text{M}}(x)$, where $f^{t}_{\chi,\text{M}}(\cdot)$ follows from (\ref{eq: thm doppler pdf}).
\end{corollary}

\begin{corollary}\normalfont
	The CDF of the maximum Doppler magnitude $F_{\chi_{\max},\text{M}}^{t}(x)$ within a cluster of $N$ UEs is given by
	$F_{\chi_{\max},\text{M}}^{t}(x)=\left(F^{t}_{\chi,\text{M}}(x)\right)^N$, where $F^{t}_{\chi,\text{M}}(\cdot)$ follows from (\ref{eq: thm doppler}). The corresponding PDF is given by
	$f_{\chi_{\max},\text{M}}^{t}(x)=N \left(F^{t}_{\chi,\text{M}}(x)\right)^{N-1}f^{t}_{\chi,\text{M}}(x)$, where $f^{t}_{\chi,\text{M}}(\cdot)$ follows from (\ref{eq: thm doppler pdf}).
	
\end{corollary}
\begin{corollary}\normalfont\label{cor: simple case}
	Let us consider the special case when the satellite is directly above the cluster center. We caution that the analytical CDF bound is quite loose for this case. Nevertheless, this corner case helps expose the role of different parameters on Doppler magnitude
	\begin{align}\label{eq: thm doppler special}
	F^{t}_{\chi,\text{M}}(x) = \frac{h^2}{\rho^2}\frac{x^2}{A^2-x^2} 
	\end{align}
	for $0\leq x \leq \frac{A}{\sqrt{1+\frac{h^2}{\rho^2}}}$, and the PDF is given by
	\begin{align}\label{eq: thm doppler special pdf}
	f^{t}_{\chi,\text{M}}(x) = \frac{2A^2 h^2}{\rho^2}\frac{x}{\left(A^2-x^2\right)^2}. 
	\end{align}
	\begin{proof}
		The results follow by plugging $\hat{R}_{t}=0$ in Theorem \ref{th: matern} and Corollary \ref{cor:pdf}.
	\end{proof}
\end{corollary}
\begin{remark}\normalfont\label{rem: tradeoff}
It follows from (\ref{eq: thm doppler special}) that the Doppler magnitude at a random UE tends to be smaller when the satellite altitude is increased or the cluster radius is decreased. With cluster radius fixed, doubling the altitude more than quadruples the probability of limiting the Doppler magnitude to a certain threshold $T$. Moreover, it is possible to tradeoff satellite altitude with cluster size to curtail the Doppler magnitude to $T$ with probability $\epsilon$. For a fixed $\epsilon$, the cluster area can be quadrupled (and a larger $N$ can be afforded) by doubling the altitude. Finally, the maximum Doppler shift seen by a random in-cluster UE is given by $x_{\max}=\frac{A}{\sqrt{1+\frac{h^2}{\rho^2}}}$ since $F^{t_0}_{\chi,\text{M}}(x_{\max})=1$.
\end{remark}
The following remark treats the case without clustering.
\begin{remark}\normalfont
Let us consider the special case where clustering is not used, i.e., the entire cell forms a cluster containing all the UEs associated with that cell. 
We set the cluster radius to the cell radius $\rho=R_{\text{cell}}$ in Theorem \ref{th: matern} to obtain expressions for this case. 
We note that this special case leads to a uniform UE distribution within the cell. This underlying UE distribution, admittedly, is not the same as that for the case with clustering. Despite this subtle difference, we believe that the analytical expression still provides useful system-level insights. We intend to provide a meticulous analytical treatment for the case without clustering in future work.
\end{remark}

\section{Numerical Results}\label{secSim}

In this section, 
we examine the impact of various system parameters on the CDF of the Doppler shift magnitude at a random UE within the cluster under consideration. The plots are obtained using Theorem \ref{th: matern}. We set $f_c=2$ GHz, $r_E=6371$ km, $\omega_E=7.27\times{10}^{-5}$ rad/sec, $\omega_s=0.0011$  rad/sec for $h=600$ km or $\omega_s=9.5809\times{10}^{-4}$ rad/sec for $h=1200$ km, and $\theta_i=0$. 

\begin{figure}[h]
	\includegraphics[width=0.95\columnwidth]{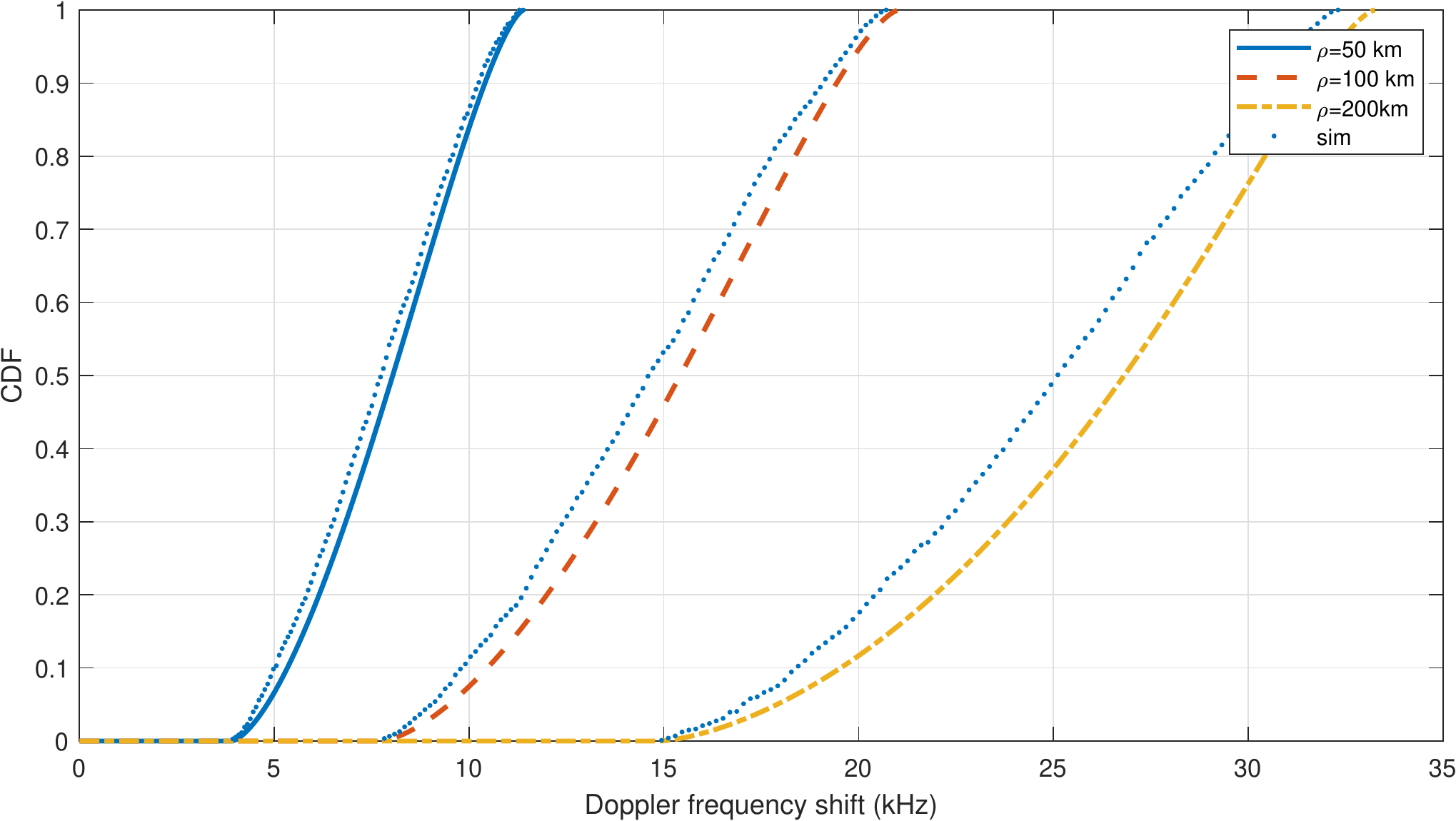}
	\caption{CDF of Doppler shift magnitude for various cluster radii for $h=600$ km. The Doppler shift gets more severe as the cluster size is increased.}
	\label{fig: cluster radius}
\end{figure}

\begin{figure}[h]
	\includegraphics[width=0.95\columnwidth]{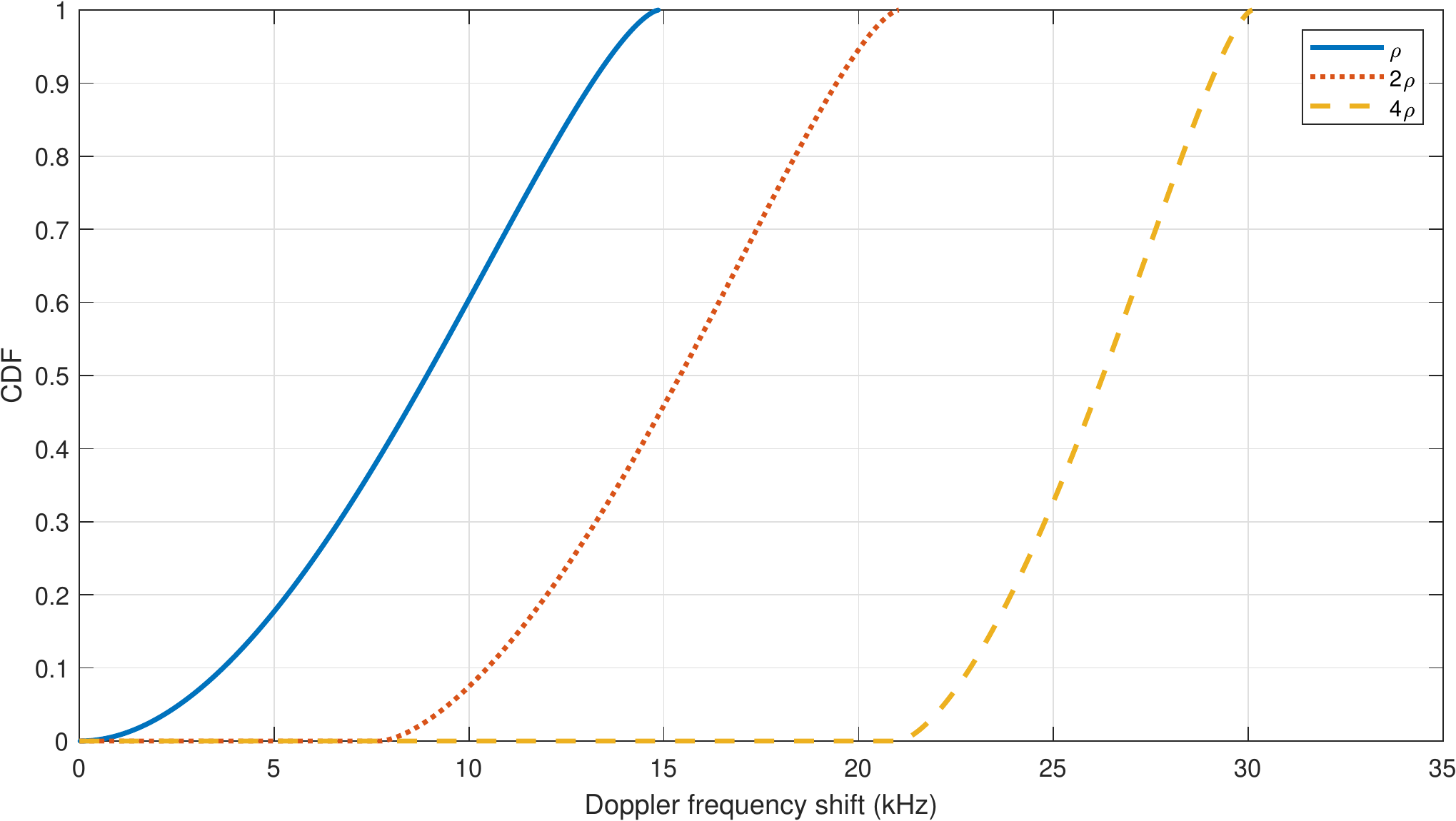}
	\caption{CDF of Doppler shift magnitude for various satellite locations for $h=600$ km and $\rho=100$ km. The CDF worsens as the satellite moves away from the cluster center.}
	\label{fig: sat location}
\end{figure}

\subsection{Impact of cluster radius}
In Fig. \ref{fig: cluster radius}, we plot the CDF of the Doppler shift magnitude for $h=600$ km at a random in-cluster UE for various values of the cluster radius $\rho$. We obtain this plot for the instant when the sub-satellite point is located at a distance $\hat{R}_{t}=2\rho$ from the cluster center and the cluster center is located on the ground track. 
We plot the CDF for two cases: i) upperbound on Doppler magnitude using Theorem \ref{th: matern}, and ii) exact Doppler magnitude obtained from Monte Carlo simulations (sim). We observe that the analytical results preserve the trend of the CDF of exact Doppler magnitude. We further note that the resulting CDF bound is tighter for smaller cluster radii. This is because the analytical framework relies on flat-earth approximation which gets more accurate for smaller distances.
Moreover, the plot reaffirms the intuition that a larger cluster suffers from a more severe Doppler shift.

\subsection{Impact of satellite location}
In Fig. \ref{fig: sat location}, we plot the Doppler CDF for various satellite locations relative to the cluster center. This is achieved by varying the distance $\hat{R}_{t}$ between the sub-satellite point and the cluster center, while fixing the satellite altitude $h=600$ km and the cluster radius $\rho=100$ km.
We recall that when $\hat{R}_{t}>\rho$, the sub-satellite point lies outside the cluster of radius $\rho$.
As the sub-satellite point moves away from the cluster center, the Doppler CDF gets worse.

\subsection{Impact of satellite altitude}
\begin{figure}[h]
	\includegraphics[width=0.95\columnwidth]{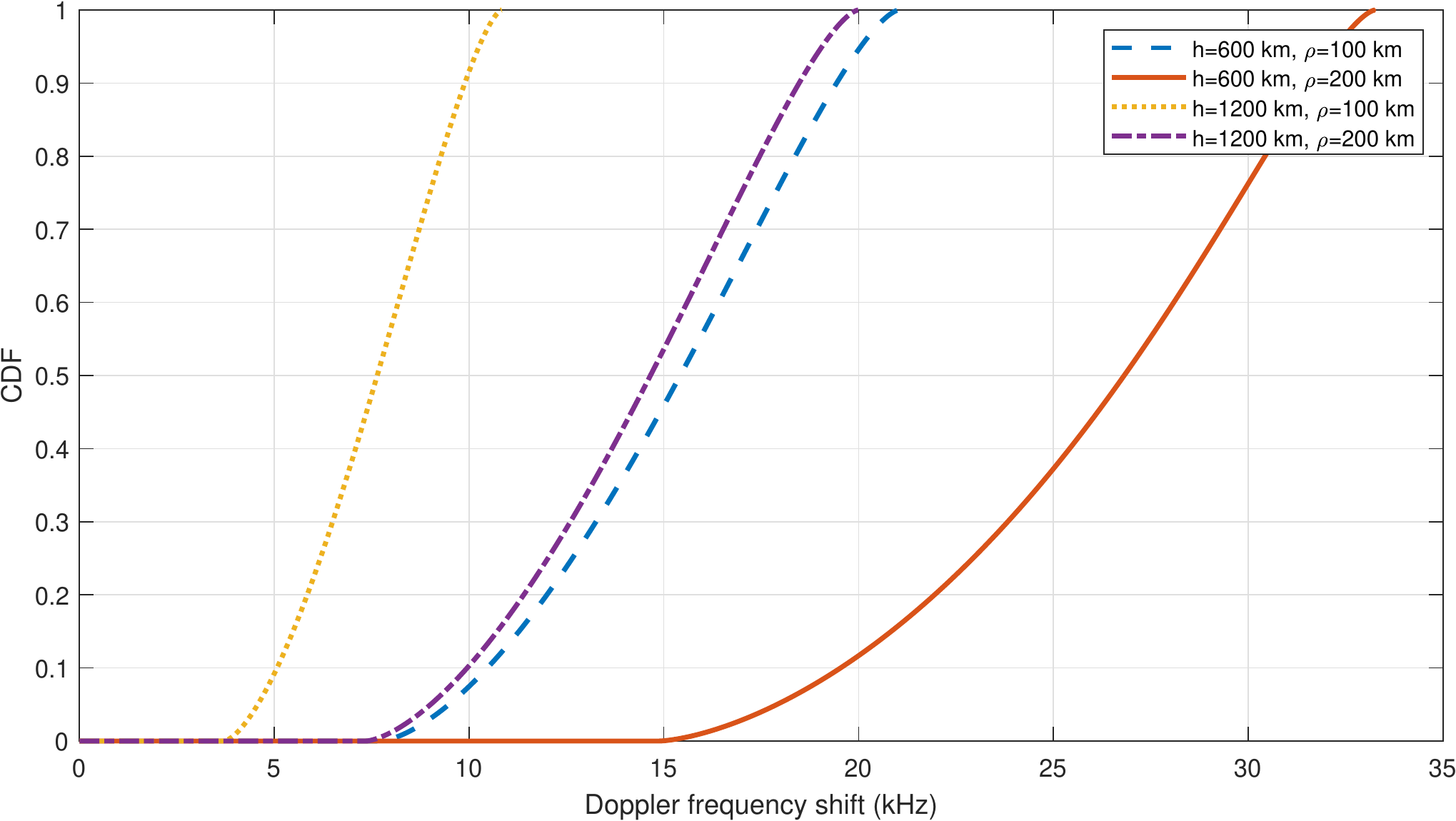}
	\caption{CDF of Doppler shift magnitude for various satellite altitudes and cluster radii. The CDF improves with an increase in satellite altitude.}
	\label{fig: height}
\end{figure}

In Fig. \ref{fig: height}, the Doppler CDF is plotted for various values of satellite altitude and cluster radius. In this plot, the sub-satellite point is assumed to be located at a distance $\hat{R}_{t}=2\rho$ from the cluster center. 
We observe that the Doppler CDF improves as the satellite altitude is increased. We further note that scaling both the satellite altitude and the cluster radius by the same factor improves the Doppler CDF. For example, consider a baseline case with a satellite altitude $h=600$ km and a cluster radius $\rho=100$ km. Doubling the satellite altitude to $h=1200$ km results in an improved CDF. If we now double the cluster radius to $\rho=200$ km, the resulting CDF still outperforms the baseline case. In other words, it is possible to support larger clusters at higher altitudes while still improving the Doppler CDF. This confirms the analytical insight obtained from Corollary \ref{cor: simple case}.

\section{Conclusions}\label{secConc}
In this paper, we characterized the Doppler magnitude distribution seen by the terrestrial UEs in a LEO satellite network. Using the framework of cluster point processes, UEs were grouped into clusters to reduce the differential Doppler. Under some simplifying assumptions, we characterized the CDF and PDF of the Doppler magnitude at a randomly located UE within a cluster for a given satellite location. The analytical and numerical results revealed several insights on the interplay between various system parameters and the Doppler shift distribution. First, the Doppler shift tends to worsen with the increase in cluster size. Second, the Doppler CDF also degrades as the satellite moves away from the cluster center. Third, the CDF improves with an increase in the satellite altitude even if the cluster size is scaled accordingly.

\balance
\bibliography{references,All_Ref_Ahmed_June,references_etal}

\begin{thebibliography}{10}

\bibitem{lin20195g}
X.~Lin, B.~Hofstr{\"o}m, E.~Wang, G.~Masini, H.-L. Maattanen, H.~Ryd{\'e}n,
  J.~Sedin, M.~Stattin, O.~Liberg, S.~Euler, {\em et~al.}, ``{5G} {New Radio}
  evolution meets satellite communications: Opportunities, challenges, and
  solutions,'' {\em arXiv preprint arXiv:1903.11219}, 2019.

\bibitem{guidotti2019architectures}
A.~{Guidotti}, A.~{Vanelli-Coralli}, M.~{Conti}, S.~{Andrenacci},
  S.~{Chatzinotas}, N.~{Maturo}, B.~{Evans}, A.~{Awoseyila}, A.~{Ugolini},
  T.~{Foggi}, L.~{Gaudio}, N.~{Alagha}, and S.~{Cioni}, ``Architectures and key
  technical challenges for {5G} systems incorporating satellites,'' {\em IEEE
  Trans. Veh. Technol.}, vol.~68, pp.~2624--2639, March 2019.

\bibitem{ganz1994performance}
A.~Ganz, Y.~Gong, and B.~Li, ``Performance study of low earth-orbit satellite
  systems,'' {\em IEEE Trans. Commun.}, vol.~42, no.~234, pp.~1866--1871, 1994.

\bibitem{ali1998doppler}
I.~Ali, N.~Al-Dhahir, and J.~E. Hershey, ``Doppler characterization for {LEO}
  satellites,'' {\em IEEE Trans. Commun.}, vol.~46, no.~3, pp.~309--313, 1998.

\bibitem{ali1997doppler}
I.~Ali, N.~Al-Dhahir, J.~E. Hershey, G.~J. Saulnier, and R.~Nelson, ``Doppler
  as a new dimension for multiple access in {LEO} satellite systems,'' {\em
  International journal of satellite communications}, vol.~15, no.~6,
  pp.~269--279, 1997.

\bibitem{li2002analytical}
S.-Y. Li and C.~Liu, ``An analytical model to predict the probability density
  function of elevation angles for {LEO} satellite systems,'' {\em IEEE Commun.
  Lett.}, vol.~6, no.~4, pp.~138--140, 2002.

\bibitem{3GPPTR}
{3GPP TR 38.821}, ``Solutions for {NR} to support non-terrestrial networks
  {(NTN)},'' 2019.

\bibitem{guidotti2017satellite}
A.~Guidotti, A.~Vanelli-Coralli, M.~Caus, J.~Bas, G.~Colavolpe, T.~Foggi,
  S.~Cioni, A.~Modenini, and D.~Tarchi, ``Satellite-enabled {LTE} systems in
  {LEO} constellations,'' in {\em Proc. 2017 IEEE Int. Conf. Commun. Workshops
  (ICC Workshops)}, pp.~876--881, May 2017.

\bibitem{maxdoppler2018NBIoT}
O.~{Kodheli}, S.~{Andrenacci}, N.~{Maturo}, S.~{Chatzinotas}, and F.~{Zimmer},
  ``Resource allocation approach for differential doppler reduction in {NB-IoT}
  over {LEO} satellite,'' in {\em Proc. 2018 9th Advanced Satellite Multimedia
  Systems Conference and the 15th Signal Processing for Space Communications
  Workshop (ASMS/SPSC)}, pp.~1--8, Sep. 2018.

\bibitem{kodheli2019journal}
O.~{Kodheli}, S.~{Andrenacci}, N.~{Maturo}, S.~{Chatzinotas}, and F.~{Zimmer},
  ``An uplink {UE} group-based scheduling technique for {5G} m{MTC} systems
  over {LEO} satellite,'' {\em IEEE Access}, vol.~7, pp.~67413--67427, 2019.

\bibitem{3GPPHetNetSahaAfsShi2018}
C.~{Saha}, M.~{Afshang}, and H.~S. {Dhillon}, ``{3GPP}-inspired {H}et{N}et
  model using {P}oisson cluster process: Sum-product functionals and downlink
  coverage,'' {\em IEEE Trans. Commun.}, vol.~66, pp.~2219--2234, May 2018.

\bibitem{pritchard1993satellite}
W.~L. Pritchard, H.~G. Suyderhoud, and R.~A. Nelson, {\em Satellite
  communication systems engineering}.
\newblock Prentice-Hall, Inc., 1993.

\bibitem{mathai1999introduction}
A.~M. Mathai, {\em An introduction to geometrical probability: distributional
  aspects with applications}, vol.~1.
\newblock CRC Press, 1999.

\end{thebibliography}
\bibliographystyle{ieeetr}
\end{document}